\documentclass[preprint,showkeys,amsmath,amssymb]{revtex4}

\usepackage{bm}

\newcommand{\al}{\alpha}
\newcommand{\be}{\beta}
\newcommand{\ga}{\gamma}
\newcommand{\de}{\delta}
\newcommand{\eps}{\varepsilon}

\newcommand{\la}{\lambda}
\newcommand{\om}{\omega}

\newcommand{\Ga}{\Gamma}

\newcommand{\Om}{\Omega}

\newcommand{\N}{\mathcal{N}}

\newcommand{\lab}{\bar{\lambda}}
\newcommand{\Gb}{\bar{G}}

\newcommand{\wb}{\bar{w}}
\newcommand{\Vb}{\bar{V}}
\newcommand{\psib}{\bar{\psi}}
\newcommand{\phib}{\bar{\phi}}
\newcommand{\Omb}{\bar{\Omega}}

\newcommand{\At}{\tilde{A}}
\newcommand{\at}{\tilde{a}}
\newcommand{\chit}{\tilde{\chi}}
\newcommand{\Ct}{\tilde{C}}
\newcommand{\ct}{\tilde{c}}
\newcommand{\tht}{\tilde{\theta}}

\newcommand{\pa}{\partial}

\newcommand{\Be}{\begin{equation}}
\newcommand{\ee}{\end{equation}}

\begin{document}


\title{On the cohomology and inner products of the Berkovits superparticle and superstring}
\author{Michael Chesterman}
 \email{Michael.Chesterman@kau.se}
\affiliation{The Physics Department, Karlstad University, S-651 88
Karlstad, Sweden.}

\date{April 2004}

\begin{abstract}
We describe the complete cohomology of the Berkovits BRST operator
for the superparticle. It is non-zero at eight ghost numbers,
splitting into two quartets, the members of each quartet being
completely isomorphic. Based only on considerations of the
isomorphisms of the cohomology, and using only the standard inner
product, we derive the inner product appropriate for string
amplitudes. It is in agreement with Berkovits' conjectured
prescription, which is one element of an equivalence class. We
discuss the Chern-Simons style action for D=10 super Yang-Mills,
which is now manifestly superspace covariant.
\end{abstract}



\maketitle

\section{Introduction} \label{sec:introduction}

The old problem of the covariant quantization of the superstring
has seen some real progress in recent years with Berkovits'
approach using pure spinors
\cite{Berkovits:2000fe,Berkovits:2000ph,Berkovits:2001rb,Berkovits:2001mx,Berkovits:2002qx}.
See ref. \cite{Berkovits:2002zk} for a review. The interest in a
$D=10$ super-Poincar\'e covariant quantization is largely due to
the hope that it can fill in gaps in the computational power of
the quantized RNS and light-cone gauge Green-Schwarz strings.
These gaps are most notably amplitudes with more than four
external fermions, covariant quantization in Ramond-Ramond
backgrounds, and higher loop amplitudes.

An unsolved problem in Berkovits' approach is to understand the
origin of his conjectured measure, which is used for on-shell
calculations such as string amplitudes, and off-shell to construct
a cubic string field theory type action for super Yang-Mills and
potentially a full open string field theory action
\cite{Berkovits:2002zk}. For simplicity, we study the
superparticle instead of the superstring wherever possible. The
measure is defined for the superparticle indirectly, by the
prescription that
\begin{equation}\label{eq:Berk_prescription}
\langle (\la\Ga^a\theta) (\la\Ga^b\theta) (\la\Ga^c\theta)
(\theta\Ga_{abc}\theta)\rangle = 1,
\end{equation}
and roughly speaking expectations of all other covariant
combinations of $\theta^\al$ and $\la^\al$ are zero, where
$\theta^\al$ are fermionic superspace co-ordinates, $\la^\al$ the
bosonic pure spinor ghosts, and $\Ga^a$ the gamma matrices.

As well as understanding where the above prescription comes from,
it is natural to ask questions, such as whether it is unique. What
explicit form should the measure $\mu(\theta,\la)$ take? Can we
construct a superspace covariant super Yang-Mills, and string
field theory action? Note that the prescription in equation
\eqref{eq:Berk_prescription} isn't manifestly superspace covariant
\cite{Berkovits:2001rb}.

We find that the clues needed to answer these questions come from
an analysis of the full superparticle cohomology. So far, the
cohomology has been studied at positive ghost number
\cite{Berkovits:2001rb,Berkovits:2002zk,Cederwall:2001dx}, where
it is believed to be non-zero at ghost numbers 8, 9, 10, 11, given
a convenient definition of the ghost number operator. There is
also however non-zero cohomology at ghost numbers $-8$, $-9$,
$-10$, $-11$. In my previous work
\cite{Chesterman:2002ey,Chesterman:2002a}, the cohomology at ghost
numbers $\pm9$ was used to calculate amplitudes for the free
superparticle with the standard inner product. However, the
cohomology as a whole was not studied. Our interest in the present
paper is mainly in the isomorphisms between the cohomologies, and
their relation to inner products.

There is some overlap between this work and the recent paper of
Grassi \textit{et al.} \cite{Grassi:2004nz}, concerning their
alternative superstring approach without using pure spinors. A
measure is derived using eqn \eqref{eq:Berk_prescription} as a
starting point, which essentially corresponds with a result
derived in this article. However, our approach is rather
different, for example we don't assume the Berkovits prescription
\textit{a priori}, but instead derive it.

The paper is organized as follows. In section \ref{sec:coho} we
study the superparticle cohomology, its isomorphisms and inner
products, we similarly study the open string cohomology in section
\ref{sec:string_coho}, we apply these results to open superstring
amplitudes in section \ref{sec:string_amps}, and to super
Yang-Mills and open string field theory in section \ref{sec:SYM}.

\section{The Complete Superparticle Cohomology} \label{sec:coho}
\subsection{The Cohomology}\label{sec:coho_subsec}
We begin with the phase space for Berkovits' $D=10$, $\N=1$
superparticle \cite{Berkovits:2001rb,Berkovits:2002zk}. It has
canonical co-ordinates given by $(x^a , \theta^\al, \la^\al)$ and
their respective conjugate momenta $(p_a, p_\al, w_\al)$, where
$a=0,\ldots 9$, $\al = 1, \ldots, 16$, $\theta^\al$ is a
Majorana-Weyl anti-commuting spinor, $\la^\al$ a Weyl commuting
complex spinor, and $(x^a, \theta^\al)$ are the usual superspace
co-ordinates. We quantize in the Schr\"odinger representation
where $(p_a, p_\al, w_\al)\equiv (-i\pa/\pa x^a,
-i\pa/\pa\theta^\al, -i\pa/\pa\la^\al)$.


The Berkovits BRST operator is given by
\begin{eqnarray}
Q=\la^\al D_\al, && Q^2 = -i\la \Ga^a \la \frac{\pa}{\pa x^a}
\end{eqnarray}
where
\begin{eqnarray}
D_\al \equiv \frac{\pa}{\pa \theta^\al} - i \Ga^a_{\al\be}
\theta^\be \frac{\pa}{\pa x^a}
\end{eqnarray}
is the usual superspace covariant derivative, and $\Ga^a_{\al\be}$
are off-diagonal components of the full $32 \times 32$ gamma
matrices in the Weyl representation.

In order that $Q$ be nilpotent, the ghosts obey pure spinor
constraints
\begin{equation}
\la^\al \Ga^a_{\al\be} \la^\be = 0,
\end{equation}
which are imposed as first class constraints. Observables are
required to be gauge invariant with respect to them. There are
essentially two approaches available to quantize the ghosts. One
is by canonical gauge-fixing using $U(5)$ co-ordinates, in a
manner analogous to choosing light-cone gauge for the bosonic
particle \cite{Chesterman:2002ey,Chesterman:2002a}. The second is
a BRST approach, in which a second BRST operator $Q_{gc} = C_a
\la\Ga^a\la + \ldots$ is introduced to form a BRST double complex
\cite{Chesterman:2002ey,Chesterman:2002a}, where $C_a$ are
fermionic ghosts with a separate ghost number grading to $\la^\al$
and $w_\al$, and the ellipses refer to ghosts-for-ghosts terms.
There is no known completion of these terms. However, for many
applications, we don't need to know them. For the sake of
covariance, and not having to constrain $\la^\al$, we use the
latter approach.

It was seen in \cite{Chesterman:2002ey,Chesterman:2002a} that all
conceivable ghosts-for-ghosts terms in $Q_{gc}$ automatically
annihilate wave functions at the two `physical' ghost numbers,
which is a typical feature of the Schr\"odinger representation.
Thus, we need only consider the first term of $Q_{gc}$. Generic
wave functions take the form $F_{\al\be
...}(x,\theta)\la^\al\la^\be ... \phi_{C=0,U=0}$ and $G^{\al\be
...}(x,\theta)w_\al w_\be ... \de^{(16)}(\la) \phi_{B=0,V=0}$,
where $C$ and $U$ refer to all the fermionic and bosonic ghosts
respectively, $B$ and $V$ to their ghost momenta, and the ghost
wave functions  are defined as $\phi_{C=0,U=0} \equiv
\de(C)\de(U)$ and $\phi_{B=0,V=0}\equiv 1$.

We define the ghost number operator
\begin{eqnarray}\label{eq:G_defn}
G = \frac{i}{2}(\la^\al w_\al + w_\al \la^\al) && G^\dag = -\Gb,
\end{eqnarray}
where $\Gb= G(\lab,\wb)$, and where $\lab^\al$ and $\wb_\al$
correspond to the complex conjugates of $\la^\al$ and $w_\al$
respectively. Differently to
\cite{Chesterman:2002a,Chesterman:2002ey}, we ignore the extra
ghost terms, preempting that we will factor out the $C$ and $U$
ghost wave function from physical states. As shown later in this
section, the generalized antihermicity condition obeyed by $G$
ensures that the state cohomology forms doublets at ghost number
$\pm g$.

The upshot of the BRST approach, after factorizing out the $C$ and
$U$ ghost wave functions for simplicity, is that there are two
different definitions of physical states, depending on the ghost
number. For $g \geq 8$, a physical state $\psi_g$ at ghost number
$g$ is defined by
\begin{eqnarray}\label{eq:psi_g_geq_8}
Q \psi_g \approx 0, && \de \psi_g \approx Q \psi_{g-1},
\end{eqnarray}
where
\begin{eqnarray}
\psi_1 \approx \psi_2 &\Rightarrow & \psi_1 = \psi_2 + \la\Ga^a\la
\phi_a
\end{eqnarray}
for some wave function $\phi_a$. Whereas for $g \leq 8$,
\begin{eqnarray}\label{eq:psi_g_leq_8_1}
\la\Ga^a\la \psi_{g} = 0, && Q \psi_{g} =0,\\
\de \psi_{g} = Q \psi_{g-1}, && \la\Ga^a\la \psi_{g-1} =
0.\label{eq:psi_g_leq_8_2}
\end{eqnarray}
Note that there are no wave functions with $-8 < g < 8$, as is
standard for bosonic ghosts. For negative ghost number, the
approach with the ghost constraints is essentially the same as
Dirac's old covariant approach.


Some typical wave functions are
\begin{eqnarray}\label{eq:ex_wavfncs}
\psi_{-10} = \om_\al \om_\be  A^{\al\be}(x,\theta)
\de^{(16)}(\la), && \psi_{11} = \la^\al\la^\be\la^\ga
\Ct_{\al\be\ga}(x,\theta).
\end{eqnarray}
The equations of motion and gauge transformation of
$A^{\al\be}(x,\theta)$ coming from \eqref{eq:psi_g_leq_8_1} and
\eqref{eq:psi_g_leq_8_2} are
\begin{eqnarray}
\Ga^{a}_{\al\be} A^{\al\be} = 0, && D_\al A^{\al\be} = 0,\\
\de A^{\al\be} = D_\ga \phi^{\al\be\ga}, && \Ga^a_{\al\be}
\phi^{\al\be\ga} = 0,
\end{eqnarray}
where $\phi^{\al\be\ga}$ is an arbitrary superfield, and all the
superfields are symmetric in their spinor indices. It is more
elegant to keep the ghosts $\la^\al$ when describing the equations
of motion and gauge transformations of superfields at positive
ghost number
\begin{eqnarray}
\la^\al \la^\be \la^\ga \la^\de D_\al \Ct_{\be\ga\de} \approx 0,
&& \la^\al \la^\be \la^\ga \de \Ct_{\al\be\ga} \approx \la^\al
\la^\be \la^\ga D_\al \phi_{\be\ga},
\end{eqnarray}
where $\phi_{\al\be}$ is an arbitrary superfield.

The complete state cohomology $H^*(Q)$ is summarised below in
table \ref{tab:cohomology}
\begin{table}[h]
\caption{The Superparticle Cohomology}\label{tab:cohomology}
\begin{tabular}{|c|c|c|c|}
\hline
Ghost number & Wave function & Physical content & $\theta$ levels\\
\hline -11 & $C^{\al\be\ga}(x,\theta)$ & $c(x)$ & 11\\
\hline
-10 & $A^{\al\be}(x,\theta)$ & $a_a(x), \chi^\al(x)$ & 12,13\\
\hline
-9  & $\At^{\al}(x,\theta)$ & $\at^a(x), \chit_\al(x)$ & 15,14\\
\hline
-8 & $\Ct(x,\theta)$ & $\ct(x)$ & 16\\
\hline
8 & $C(x,\theta)$ & $c(x)$ & 0\\
\hline
9 & $A_{\al}(x,\theta)$ & $a_a(x), \chi^\al(x)$ & 1,2\\
\hline
10 & $\At_{\al\be}(x,\theta)$ & $\at^a(x),\chit_\al(x)$ & 4,3\\
\hline
11 & $\Ct_{\al\be\ga}(x,\theta)$ & $\ct(x)$ & 5\\
\hline
\end{tabular}
\end{table}

where
\begin{eqnarray}\label{eq:phys_fields1}
\pa_a c(x) = 0, && \de \ct(x) = \pa_a b^a(x),\\
\pa^a (\pa_a a_b - \pa_b a_a) = 0, && \de \at^a = \pa^b( \pa_a
s_b(x) - \pa_b s_a(x)),\\
\pa_a \at^a = 0, &&  \de a_a(x) = \pa_a \phi(x),\\
\Ga^a_{\al\be} \pa_a \chi^\be = 0, && \de \chit_\al =
\Ga^a_{\al\be} \pa_a\phi^\be(x),\label{eq:phys_fields4}
\end{eqnarray}
and where $b^a, s_a, \phi, \phi^\be$ are arbitrary functions of
$x$. The cohomology at all other ghost numbers is null. We show
only the superfield part of the wave functions, since the ghost
part can be easily deduced from eqn \eqref{eq:ex_wavfncs}. The
fourth column for $\theta$ level tells us at what power of
$\theta$ the physical content lies. The physical content of the
cohomology is given by two copies of the on-shell fields of
super-Maxwell theory in the Batalin-Vilkovisky formalism. These
are the photon, photino and ghost fields $a_a(x)$, $\chi^\al(x)$,
$c(x)$ and their respective anti-fields $\at^a(x)$, $\chit_\al(x)$
and $\ct(x)$.

The positive ghost number cohomology was first deduced in
\cite{Berkovits:2001rb,Berkovits:2002zk}, including a proof that
$H^g(Q)$ is null for $g>11$. Also, a direct computation of the
representation content of this cohomology was made
\cite{Cederwall:2001dx} using the computer program LiE
\cite{Cohen:1998}, which agrees with Berkovits' results. Below is
given a convenient representative of each cohomology class at
positive ghost number
\begin{eqnarray}\label{eq:psi_g_1}
C(x,\theta) &=& c,\\
\la^\al A_\al(x,\theta) &\sim& a_a(x) \la \Ga^a \theta +
(\chi(x)\Ga_{abc} \la) (\theta\Ga^{abc} \theta) + \ldots,\label{eq:A_al}\\
\la^\al \la^\be \At_{\al\be}(x,\theta) &\sim& \la\Ga^{abcde}\la[
(\theta\Ga_{abc}\theta) (\theta \Ga_{de})^\al \chit_\al(x) +
(\theta\Ga_{abc}\theta)(\theta \Ga_{def}\theta)\at^f(x)  + \ldots
] \label{eq:At_albe}\\
\la^\al\la^\be\la^\ga \Ct_{\al\be\ga}(x,\theta) &\sim& \ct
(\la\Ga^a\theta) (\la\Ga^b\theta) (\la\Ga^c\theta)
(\theta\Ga_{abc}\theta),\label{eq:psi_g_4}
\end{eqnarray}
where $c$ and $\ct$ are constants, and $\psi \sim \psi'
\Rightarrow \psi = \psi' + Q \phi$ for some wave function $\phi$.
We argue later in section \ref{sec:iso_innprod} that we can always
choose a Wess-Zumino type gauge such that the superfields $A_\al$
and $\At_{\al\be}$ terminate at $\theta$-level five.

The negative ghost number cohomology follows from the above since
the cohomology $H^{-g}(Q)$ is isomorphic to $H^g(Q)$ for each $g$,
as a result of the usual non-degenerate, well-defined inner
product. For a discussion, see
\cite{Henneaux:1992ig,Chesterman:2002ey,Chesterman:2002a}. Roughly
speaking, a basis of BRST-closed but not exact states
$\{\psi^A_g\}$ can be chosen, such that
\begin{equation}
\langle\psib_{g}^B|\psi_{g'}^A\rangle = \de^{AB}\de_{g+g'},
\end{equation}
where $\psib_g = \psi_g(\lab,x,\theta)$ and $\{\psi^A_g\}$ are
orthogonal to all other states. So $\psi^A_g$ maps to
$\psib^A_{-g}$ under the isomorphism. As in
\cite{Chesterman:2002ey,Chesterman:2002a}, we place barred states
on the left in the inner product. One can see that states at ghost
number $g$ couple to states at ghost number $-g$ for finite inner
product, since by considering $\langle\psib_{g}|G|
\psi_{g'}\rangle$ with eqn \eqref{eq:G_defn}, we find
\begin{equation}
(g+g')\langle\psib_{g}| \psi_{g'}\rangle = 0.
\end{equation}
The general prescription to calculate $H^{-g}(Q)$ given $H^g(Q)$
is that a field at $\theta$-level $l$ in $\psi_g$ becomes its
anti-field at $\theta$-level $16-l$ in $\psi_{-g}$. Note that
under the inner product, $(\theta)^l$ couples to $(\theta)^{16-l}$
due to the integration $\int{d^{16}\theta}$ . So $\theta^\al$ is
replaced by $\pa/\pa\theta^\al$ and $\la^\al$ by $\pa/\pa\la^\al$,
with a $\de^{16}(\theta)\de^{16}(\la)$ placed at the end, where
$\de^{(16)}\theta \equiv \prod_{\al} \theta^\al$. As an example
\begin{eqnarray}\label{eq:psi_minus_11}
\psi_{-11} \sim c (\om \Ga^a \tht) (\om \Ga^b \tht) (\om \Ga^c
\tht) (\tht \Ga_{abc}\tht) \de^{(16)}(\la),
\end{eqnarray}
where $c \in \mathbb{C}$ is constant and
\begin{equation}
\tht_\al \tht_\be ... \equiv \frac{\pa}{\pa \theta^\al}
\frac{\pa}{\pa \theta^\be} ... \de^{(16)}(\theta).
\end{equation}

\subsection{Isomorphisms and inner products}\label{sec:iso_innprod}
We have discussed the isomorphisms between $H^{g}(Q)$ and
$H^{-g}(Q)$. However, there are further isomorphisms, which are
perhaps unexpected from a BRST point of view, since they are
specific to this system. From table \ref{tab:cohomology}, we see
that there are only two independent cohomologies $H^8(Q)$ and
$H^9(Q)$. All others are isomorphic to one or the other.
Specifically
\begin{eqnarray}\label{eq:iso_coho_8}
H^8(Q)\cong H^{-8}(Q) \cong H^{11}(Q) \cong H^{-11}(Q) \cong
\mathbb{C},\\
H^9(Q) \cong H^{-9}(Q) \cong H^{10}(Q) \cong H^{-10}(Q) \cong
\mathbf{8}+\mathbf{8}, \label{eq:iso_coho_9}
\end{eqnarray}
where $\mathbf{8}+\mathbf{8}$ refers to the on-shell degrees of
freedom of super Yang-Mills.

Since the isomorphism $H^g(Q)\cong H^{-g}(Q)$ is associated with
the standard inner product, we might expect to find other inner
products associated with other isomorphisms. This is indeed the
case. However, rather than looking for such inner products, we
look for explicit one-to-one maps between cohomologies and use the
standard inner product. For example, to find the inner product
relating $H^9(Q)$ to $H^{10}(Q)$, we first look for a map $f:
H^9(Q) \to H^{-10}(Q)$, then $H^{-10}(Q)$ couples to $H^{10}(Q)$
under the standard inner product.

In order to find such a mapping, first note that the BRST
invariant wave function $\la^\al A_\al$ is also a BRST invariant
operator, since $Q \,\la^\al A_\al \equiv [Q, \la^\al A_\al]$. The
same is true for all wave functions with $g\geq 8$. One can
therefore think of $\psi_9$ as the BRST invariant operator
$\la^\al A_\al$ acting on the BRST invariant vacuum state
$\psi_8(1)\equiv 1$
\begin{equation}
\psi_9(a_a,\chi^\al) = \la^\al A_\al(a_a,\chi^\al) \psi_8(1),
\end{equation}
where we write BRST-closed states $\psi_g$ as functions of their
physical content, as described in eqns \eqref{eq:psi_g_1} to
\eqref{eq:psi_g_4}. Other BRST-closed states for $g\geq 8$ are
built in a similar manner.

We can now construct different BRST-invariant states by replacing
the vacuum state. States in any of the cohomologies isomorphic to
$H^8(Q)$ in eqn \eqref{eq:iso_coho_8} are candidates. We choose
$\psi_{-11}(1)$, where $\psi_{-11}$ is given in equation
\eqref{eq:psi_minus_11}. The claim is then that
\begin{eqnarray}\label{eq:psi_map_1}
\psi_{-11}(c) \sim C(c) \psi_{-11}(1), && \psi_{-10}(a_a,
\chi^\al) \sim
\la^\al A_\al(a_a,\chi^\al) \psi_{-11}(1),\\
\psi_{-9}(\at^a, \chit_\al) \sim \la^\al\la^\be
\At_{\al\be}(\at^a, \chit_\al) \psi_{-11}(1), && \psi_{-8}(\ct)
\sim \la^\al\la^\be\la^\ga \Ct_{\al\be\ga}(\ct) \psi_{-11}(1),
\label{eq:psi_map_2}
\end{eqnarray}
at least up to a rescaling of the physical fields, which can be
taken care of by calculating suitable factors in the superfield
expansions of eqns \eqref{eq:psi_g_1} to \eqref{eq:psi_g_4}. This
is reminiscent of the mapping $c: H^{-1/2}(Q) \to H^{1/2}(Q)$ for
the bosonic particle, where $c$ is the worldline
reparameterization ghost.

A direct calculation of the above is feasible but time-consuming,
except in the case of $\psi_{-11}(c)$. However, we can see why it
must be true indirectly. Consider the expression for $\psi_{-10}$
in \eqref{eq:psi_map_1}. From the point of view of the physical
content of $\psi_{-10}$, the scalar $c$ in $\psi_{-11}(c)$ is at
$\theta$-level 11, and the vector and spinor $a_a$ and $\chi^\al$
are at levels one and two respectively in $\la^\al A_\al$. This
leads to $a_a$ and $\chi^\al$ being at levels 12 and 13 in
$\psi_{-10}$, which is consistent with table \ref{tab:cohomology}.
Given that $\la^\al A_\al$ and $\psi_{-11}$ are BRST-closed but
not BRST-exact, the same must be the case for $\psi_{-10}$, and
hence the physical fields $a_a$ and $\chi^\al$ in $\la^\al A_\al$
must map to the $a_a$ and $\chi^\al$ in $\psi_{-10}$ since there
are no other candidate physical fields. Similar arguments apply at
the other ghost numbers.

We now have a non-degenerate inner product between the
cohomologies $H^g(Q)$ and $H^{13-g}(Q)$ for $8 \leq g \leq 11$,
which will be useful later for string amplitude and string field
theory calculations. Defining the vacuum states
\begin{eqnarray}
|0\rangle \equiv 1, && |\Om\rangle \sim \psi_{-11}(1),
\end{eqnarray}
the inner product between states/operators $\psi$ and $\phi$ is
written
\begin{equation}\label{eq:inner_prod}
\langle\phib \Omb|\psi\rangle = \langle\Omb|\phib^\dag
\psi|0\rangle.
\end{equation}
In the case where the $(x,\theta)$ part of the wave function
$\phi$ is real, $\phib^\dag = \phi$. The vacuum states are
supersymmetric in the sense that
\begin{eqnarray}
Q_\al |0\rangle = 0, && Q_\al |\Om\rangle \sim
0,\label{eq:Om_is_Q_al_invar}
\end{eqnarray}
where $Q_\al$ is the supersymmetry generator
\begin{eqnarray}
Q_\al \equiv \frac{\pa}{\pa \theta^\al} + i \Ga^a_{\al\be}
\theta^\be \frac{\pa}{\pa x^a}, && \{Q_\al, Q \} = 0.
\end{eqnarray}
We know that $Q_\al |\Om>$ is BRST-exact because it is
BRST-closed, and using the representative $\psi_{-11}(1)$ in eqn
\eqref{eq:psi_minus_11} for $|\Om\rangle$, must consist of only
one term at $\theta$-level 10, which does not correspond to the
only non-exact but closed state in \eqref{eq:psi_minus_11}.

As an example, the inner product between the BRST closed states
$\psi_9$ and $\psi_{10}$ is
\begin{eqnarray}
\langle\Omb|\la^\al \la^\be \At_{\al\be}^\dag \la^\ga A_\ga
|0\rangle = \langle\psi_{-9}(\at^a,\chit_\al)|
\psi_{9}(a_a,\chi^\al)\rangle \propto \int{d^{10}x}((\at^a)^* a_a
+ (\chit_\al)^* \chi^\al),
\end{eqnarray}
where the last equality is explained in
\cite{Chesterman:2002ey,Chesterman:2002a}. Note that this
expression is invariant under $\de \Om = Q \psi_{-12}$, as well as
possessing the usual BRST symmetries and supersymmetry with
respect to $\la^\al \la^\be \At_{\al\be}$ and $\la^\ga A_\ga $. In
the case where we specifically take the representative for
$|\Om\rangle$ in eqn \eqref{eq:psi_minus_11}, $\langle\Omb|$
couples only to the BRST-closed term in eqn \eqref{eq:psi_g_4}
under the standard inner product, which agrees precisely with
Berkovits' prescription in \eqref{eq:Berk_prescription}. In fact
this prescription is one element of an equivalence class.
Replacing $|\Om\rangle$ with $|\Om\rangle + Q|\phi\rangle$ for
general state $|\phi\rangle$, we obtain the most general
prescription.

An interesting observation which follows from the mapping in
\eqref{eq:psi_map_1} and \eqref{eq:psi_map_2}, is that given the
explicit choice for $\psi_{-11}(1)$ in eqn \eqref{eq:psi_minus_11}
which has $11$ $\theta$'s, all the wave functions $\psi_g$,
$g=-8,...,-11$ terminate at $\tht$-level 5. This means that
through the isomorphism $H^{-g}(Q)\cong H^{g}(Q)$, a cohomology
class representative $\psi_g$, $g=8,...,11$ can always be chosen
which terminates at $\theta$-level 5. So it is quite feasible to
calculate the few remaining terms of $A_\al$ and $\At_{\al\be}$ in
eqns \eqref{eq:A_al} and \eqref{eq:At_albe}.

It is in principle possible to calculate inner products associated
with the remaining isomorphisms. In particular, following an
approach analogous to above, we may construct an inner product
which couples $H^g(Q)$ to $H^{-19-g}(Q)$ for $g = -8,...,-11$. It
is a little more complicated however, since the BRST-closed wave
function $\om_\al \At^{\al}(p_a,p_\al)$ in the momentum picture
isn't also a BRST-closed operator. Any analogous inner product
between $H^{9}(Q)$ and $H^{-10}(Q)$ would not be Lorentz covariant
because we cannot form a Lorentz scalar from two $\chi^\al$'s.


\section{On the superstring cohomology}\label{sec:string_coho}
The above ideas extend fairly easily to the case of the
superstring, where the superparticle degrees of freedom form its
zero modes. We conjecture that the complete open string cohomology
takes the same form as the superparticle cohomology, in the sense
that cohomology at the different ghost numbers play exactly the
same roles. For example $H^8(Q)$ will contain all on-shell
space-time Batalin-Vilkovisky ghosts, $H^{11}(Q)$ all the
anti-ghosts etc., where now $Q = \oint{dz i\la^\al d_\al}$, with
$d_\al=p_\al - i(\pa X_a -
\frac{i}{2}\theta\Ga_a\theta)(\Ga^a\theta)_\al$. This is a natural
conjecture to make in view of the strong analogy noted in
\cite{Berkovits:2001rb}, between the Berkovits string and
topological strings \cite{Witten:1992fb}, which have precisely
this cohomology structure.
However, except for the ghost number one vertex operator
cohomology $H^1_{\text{op}}(Q) \equiv  H^{9}(Q)$ which is known to
give the correct string spectrum
\cite{Berkovits:2000nn,Berkovits:2002qx}, no direct calculations
of cohomology at non-zero excitation level have been made. By
reversing the logic of the previous subsection, we nevertheless
prove that the isomorphisms of eqns \eqref{eq:iso_coho_8} and
\eqref{eq:iso_coho_9} also hold for the superstring.



The generalization of the relation \eqref{eq:psi_map_1} for the
state mapping $\psi_{9} \to \psi_{-10}$ is given by
\begin{eqnarray}
\psi_9 = V(0)|0\rangle && \psi_{-10} \sim V(0) \Om(0)|0\rangle,
\end{eqnarray}
where $V(z)\in H^1_{\text{op}}(Q)$ is a ghost number one
BRST-closed unintegrated vertex operator, which is $\la^\al A_\al$
for the open string ground state,
\begin{eqnarray}\label{eq:Om_z}
\Om(z) \sim \frac{1}{(2\pi)^{16}}\int{d^{16}u(u \Ga^a \tht) (u
\Ga^b \tht) (u \Ga^c \tht) (\tht \Ga_{abc}\tht)
e^{iu_\al\la^\al}},
\end{eqnarray}
and $\tht$, $\la$ are as usual all holomorphic functions of $z$.
The vacuum $|0\rangle$ is the usual one associated with $1$ in the
state-operator correspondence. It is the direct product of the
wave function $\psi_8(1)=1$ for the zero modes, which contributes
ghost number $8$, with the Fock space vacuum for the non-zero
modes, which contributes ghost number $0$. Note also that the
ghost number $-11$ vacuum state $|\Om\rangle\equiv
\Om(0)|0\rangle$ corresponds to a direct product of the wave
function $\psi_{-11}(1)$ in eqn \eqref{eq:psi_minus_11} for the
zero modes, with the Fock space vacuum for the non-zero modes.

Since the string cohomologies have infinite size, we must prove
that the isomorphisms hold true at each level of excitation $n\geq
0$. A linearly independent basis of BRST-closed but not BRST-exact
vertex operators $V(z)$ at excitation level $n$ must map to a
basis of states $\psi_{-10}$ with these same properties, since
$|\Om\rangle$ is a BRST-closed but not BRST-exact state at level
$0$. Thus, $H^9_n(Q)\subseteq H^{-10}_n(Q)$, where $H^*_n(Q)$ is
the subcohomology with excitation level $n$. Using similar
arguments for the mapping $\psi_{10} \to \psi_{-9}$, we find also
that $H^{10}_n(Q)\subseteq H^{-9}_n(Q)$. Now we know that
$H^{9}_n(Q)\cong H^{-9}_n(Q)$ and $H^{10}_n(Q)\cong H^{-10}_n(Q)$,
due to the standard non-degenerate inner product as discussed in
section \ref{sec:coho}. Thus, we conclude
\begin{equation}
H^9(Q) \cong H^{-9}(Q) \cong H^{10}(Q) \cong H^{-10}(Q),
\end{equation}
as expected. A similar argument can be used to show that
\begin{equation}
H^8(Q)\cong H^{-8}(Q) \cong H^{11}(Q) \cong H^{-11}(Q).
\end{equation}

\section{Tree-level Open Superstring Amplitudes}\label{sec:string_amps}
For free string calculations, we may use either the standard inner
product, or a generalization of the inner product in
\eqref{eq:inner_prod} $\langle\Omb|\Vb^\dag_1(0)V_2(0)|0\rangle$,
where $|\Om\rangle \equiv \Om(0)|0\rangle$ is the ghost number
$-11$ vacuum state, and $V_i(z)$ are unintegrated vertex
operators.

The expression for tree-level amplitudes is a modification of
Berkovits' expression \cite{Berkovits:2002zk}, though now just
using the ordinary inner product between two different vacuum
states
\begin{equation}
A = \text{tr } \langle\Omb| V_1(z_1) V_2(z_2) V_3(z_3) U_4 ...
U_n|0\rangle,
\end{equation}
where $U_i = \oint{dz u_i(z)}$ are integrated vertex operators,
which are related to their unintegrated couterparts as $[Q ,u_i] =
\pa_z V_i$. Note that the ghost numbers of the two vacua sum to
$-3$, which is an important check. Similarly to the superparticle
analysis in section \ref{sec:iso_innprod}, given the cohomology
class representative for $\Om(z)$ in eqn \eqref{eq:Om_z}, the
above expression corresponds exactly with Berkovits' prescription
in equation \eqref{eq:Berk_prescription}. It is invariant under
$\de |\Om\rangle = Q |\psi_{-12}\rangle$ for arbitrary
$|\psi_{-12}\rangle$, so that Berkovits' prescription is one of a
BRST equivalence class.

All symmetries are manifest in the above expression. $A$ is
manifestly BRST invariant. It is also manifestly invariant under
superspace translations, since in particular as a generalization
of \eqref{eq:Om_is_Q_al_invar}, the vacuum states obey
\begin{eqnarray}
\oint{dz \,q_\al(z)} |0\rangle = 0, && \oint{dz\, q_\al(z)}
|\Om\rangle \sim 0,
\end{eqnarray}
where $q_\al = p_\al + i (\pa X_a - \frac{i}{6} \theta \Ga_a
\pa\theta )\Ga^a_{\al\be} \theta^\be$ and $\oint{dz q_\al(z)}$ is
the supersymmetry generator. Furthermore it is $SL(2,\mathbb{R})$
covariant since
\begin{eqnarray}\label{eq:SL2R_inv_vac}
L_{0,\pm 1}|0\rangle=0, && L_{0,\pm 1} |\Om\rangle \sim 0,
\end{eqnarray}
and the unintegrated vertex operators $V_i(z_i)$ have zero
conformal weight. The final statement of \eqref{eq:SL2R_inv_vac}
requires a proof, but it is similar to the result that $P_a P^a
\psi_9 \sim 0$ for BRST-closed $\psi_9$ in the case of the
superparticle as shown in
\cite{Chesterman:2002a,Chesterman:2002ey}.

There is a complication that massive states in general have vertex
operators with terms which need a BRST extension with respect to
the ghost constraint BRST operator $Q_{gc}= C_a
\la\Ga^a\la+\ldots$, like $\la^\al w_\al$. This is a problem,
because there is currently no known completion of the ghost terms.
However, the calculation may still be doable using methods in
\cite{Chesterman:2002ey,Chesterman:2002a}. Further investigation
is warranted. Finally, it seems reasonable that amplitudes
involving closed strings will be a straightforward generalization
of this.

\section{A Superspace Action for $D=10$ Super Yang-Mills}\label{sec:SYM}

We now apply the technique of the previous section to Berkovits'
Chern-Simons style action for super Yang-Mills. The result is not
as nice as for on-shell calculations like string amplitudes,
essentially because while the inner product described in eqn
\eqref{eq:inner_prod} is non-degenerate on-shell, it is highly
degenerate off-shell. This is clear since any state with more than
$5$ $\theta$'s is orthogonal to all other states, taking the
representative for $|\Om\rangle$ in eqn \eqref{eq:psi_minus_11}.

That being said, the BRST invariant action is
\begin{equation}\label{eq:S_SYM}
S=\text{tr}\frac{1}{2}\langle\Omb|\psi Q \psi|0\rangle +
\text{tr}\frac{g}{3} \langle\Omb|\psi \psi \psi|0\rangle
\end{equation}
where
\begin{equation}
\psi=C(x,\theta)+\la^\al A_\al(x,\theta)
+\la^\al\la^\be\At_{\al\be}(x,\theta)+\la^\al\la^\be\la^\ga
\Ct_{\al\be\ga}(x,\theta),
\end{equation}
and the Lie algebra indices have been suppressed. The equations of
motion are
\begin{equation}\label{eq:SYM_eom}
(Q \psi + g\psi .\psi)|\Om\rangle =0,
\end{equation}
which is a much less restrictive equation than the actual equation
for super Yang-Mills $Q\psi+g\psi.\psi=0$. Unlike the string
amplitude calculation, the above action and equations of motion
vary under $\de |\Om\rangle = Q |\psi_{-12}\rangle$. It was argued
in \cite{Berkovits:2001rb}, using Berkovits' prescription which
corresponds with choosing the exact representative for
$|\Om\rangle= \psi_{-11}(1)$ given in eqn \eqref{eq:psi_minus_11},
that after integrating out $\theta^\al$ and $\la^\al$, the action
\eqref{eq:S_SYM} is the standard BV one except with no auxiliary
field terms. Therefore, the equations of motion \eqref{eq:SYM_eom}
must imply no conditions on the auxiliary fields, effectively
meaning that all the auxiliary fields have become pure gauge. By
making a different choice of $|\Om\rangle$, the action is modified
and auxiliary field terms appear. The precise implications of this
are yet to be understood. Another way to understand this issue, is
that one can write
\begin{equation}
\de S = \langle\de\psib \Omb| (Q\psi+g\psi.\psi)|0\rangle=0,
\end{equation}
for every $\de\psi$. Consider the case $\de\psi = \la^\al \de
A_\al$. The mapping $\de A_\al \to \de A^{\al\be}$ given by
$\de\psi \to \de\psi|\Om\rangle$ is one-to-one on-shell, but
off-shell only a fraction of the functional space spanned by $\de
A^{\al\be}$ is covered by the map, largely because the
representation content of $A_\al$ is much smaller than the
representation content of $A^{\al\be}$. Thus, $\langle\de\psib
\Omb|$ covers only a fraction of the ghost number $-10$ state
space.

Integrating out the ghosts, the action becomes
\begin{equation}
S = \text{tr }\int{d^{10}x d^{16}\theta\, C^{\al\be\ga}(1)
(\frac{1}{2}A_\al D_\be A_\ga + \frac{g}{3} A_\al A_\be A_\ga  +
\text{BV ghost terms})},
\end{equation}
where
\begin{equation}
C^{\al\be\ga}(1) \sim (\Ga^a \tht)^\al (\Ga^b \tht)^\be (\Ga^c
\tht)^\ga (\tht \Ga_{abc}\tht)
\end{equation}
is a non-dynamical superfield. The main advantage of this
expression over that of Berkovits is that the action is now
superspace covariant. Under supersymmetry transformations we allow
$C^{\al\be\ga}$ to transform just as the other superfields
$\de_{\eps} C^{\al\be\ga} = [\eps^\rho Q_\rho, C^{\al\be\ga}]$,
where $\eps^\rho$ is a Grassmann parameter. The superspace
transformation of $C^{\al\be\ga}$ cancels auxiliary field terms in
the variation of physical fields. For example $\de_\eps a_a =
(i\eps\Ga_a\chi + \text{aux. field terms})$, where these auxiliary
field terms vanish on-shell. Recall that the action $S$ is
independent of auxiliary fields for suitable choice of
$C^{\al\be\ga}$. While the action is superspace covariant, it has
the unusual features of a non-dynamical superfield $C^{\al\be\ga}$
and missing auxiliary field terms. It will be interesting to see
how it can be applied in a string field theory context. In
particular, it is desirable to understand better the role of
$C^{\al\be\ga}$.

The full cubic string field theory action is a fairly
straightforward generalization of the super Yang-Mills action
\begin{equation}
S = \text{tr}\langle\Omb|(\frac{1}{2}V Q V + \frac{g}{3}
VVV)|0\rangle,
\end{equation}
where $V$ is the sum of ghost number 0,1,2 and 3 vertex operators.

\section{Concluding Remarks}\label{sec:conclusions}

It is interesting to note that the fact that an inner product
exists which is suitable for string amplitudes is a result of the
isomorphism structure of the cohomology. In particular, that we
can find two suitable vacuum states whose ghost number sums to
$-3$ suggests that somehow the cohomology structure of $Q$ `knows'
about string tree-level amplitudes.

\begin{acknowledgments}
I would like to thank Anders Westerberg for many helpful
discussions and for reading the manuscript.
\end{acknowledgments}



\bibliography{supers1}
\end{document}